\documentclass{article}

\usepackage{arxiv}

\usepackage[utf8]{inputenc} % allow utf-8 input
\usepackage[T1]{fontenc}    % use 8-bit T1 fonts
\usepackage{hyperref}       % hyperlinks
\usepackage{url}            % simple URL typesetting
\usepackage{booktabs}       % professional-quality tables
\usepackage{amsfonts}       % blackboard math symbols
\usepackage{nicefrac}       % compact symbols for 1/2, etc.
\usepackage{microtype}      % microtypography
\usepackage{lipsum}
\usepackage{graphicx}
\usepackage{multirow}
\usepackage{amsmath}
\usepackage{tabularx}
\usepackage{array}
\usepackage{ragged2e}
\title{Struggle First, Prompt Later: How Task Complexity Shapes Learning with GenAI-Assisted Pretesting}

\author{
  Mahir Akgun \\
  College of Information Sciences and Technology\\
  Pennsylvania State University\\
  University Park, PA 16827 \\
  \texttt{makgun@psu.edu} \\
   \And
 Sacip Toker \\
  Information Systems Engineering\\
  Atılım University\\
  Ankara, Türkiye \\
  \texttt{sacip.toker@atilim.edu.tr} 
  \\
}

\begin{document}
\maketitle
\begin{abstract}
This study explores the role of AI-assisted pretesting in enhancing learning outcomes, particularly when integrated with conversational AI tools like ChatGPT. While traditional pretesting has been shown to improve retention by activating prior knowledge, the adaptability and interactivity of AI-assisted pretesting offer new possibilities for learning in digital environments. Across three experimental studies, we investigated how pretesting strategies, task characteristics, and student motivation influence learning outcomes.

Study 1 examined the impact of pretesting in an AI-supported learning environment, comparing a pretest condition (where participants attempted a problem before consulting ChatGPT) to a no-pretest condition (where participants accessed ChatGPT immediately). Study 2 extended this investigation by comparing fixed pretesting (pre-written static questions) with AI-assisted adaptive pretesting (dynamically generated AI-driven questions) to determine whether personalized, interactive questioning enhances learning. Study 3 further explored how task complexity moderates the effectiveness of AI-assisted pretesting, shifting from a linear, well-structured analytical task to an iterative, exploratory evaluation task while also assessing the motivational dimensions of engagement.

Findings suggest that AI-assisted pretesting enhances learning outcomes, particularly when tasks require higher-order thinking and decision-making. Adaptive AI-driven pretesting was more engaging but did not necessarily improve all motivational dimensions. The results indicate that task structure plays a critical role in determining pretesting effectiveness, with AI-assisted approaches showing greater benefits in complex, exploratory tasks. These findings contribute to our understanding of how AI can optimize learning strategies, offering insights into the integration of pretesting with generative AI tools for improved cognitive and motivational outcomes. 
\end{abstract}

% keywords can be removed

\section{Introduction}
In the digital age, where information is readily accessible with a single search, the role of search engines in facilitating learning experiences has become increasingly prominent. Traditional approaches to studying often involve passive consumption of information, but recent research has suggested that actively engaging with the material through methods such as pretesting—where participants are asked to retrieve information from memory before studying it—could significantly enhance learning outcomes. Pretesting engages participants in active retrieval processes, not only assessing current knowledge but also priming the memory for subsequent encoding of related information. In essence, pretesting serves as an initial assessment or diagnostic tool before formal learning occurs and can strengthen memory traces and facilitate long-term retention.

Leveraging the ubiquitous nature of search engines, particularly Google, researchers have explored the potential benefits of pretesting in improving subsequent recall for searched-for content. While search engines provide instant access to vast amounts of information, questions have arisen regarding the depth of understanding and retention achieved through such methods. In a study by Giebl et al. \cite{giebl2021answer}, some but not all of the participants (the pretest vs. no-pretest groups) were asked to solve a challenging problem before consulting Google for needed information. The results revealed that this pretesting method significantly enhanced participants' subsequent recall of the searched content. These findings highlighted the potential of pretesting as a search-for-learning strategy to deepen learning and improve memory retention. 

Building upon these findings, the present study aims to extend the investigation into the efficacy of pretesting as a learning strategy by using generative AI (GenAI) as the information retrieval tool. Specifically, we seek to examine whether pretesting, when implemented with a GenAI model, such as ChatGPT, can yield similar enhancements in participants' subsequent recall of the searched content. By exploring the potential of ChatGPT as a learning aid, we aim to contribute to a deeper understanding of optimal learning strategies in the age of AI.

In the digital age, where information is readily accessible with a single search, the role of search engines in facilitating learning experiences has become increasingly prominent. Traditional approaches to studying often involve passive consumption of information, but recent research has suggested that actively engaging with the material through methods such as pretesting—where participants are asked to retrieve information from memory before studying it—could significantly enhance learning outcomes. Pretesting engages participants in active retrieval processes, not only assessing current knowledge but also priming the memory for subsequent encoding of related information. In essence, pretesting serves as an initial assessment or diagnostic tool before formal learning occurs and can strengthen memory traces and facilitate long-term retention.

Leveraging the ubiquitous nature of search engines, particularly Google, researchers have explored the benefits of pretesting in improving subsequent recall for searched-for content. The potential of pretesting, as a search-for-learning strategy, in improving memory retention has been underscored in the literature. % While search engines provide instant access to vast amounts of information, questions have arisen regarding the depth of understanding and retention achieved through such methods. In a study by Giebl et al. \cite{giebl2021answer}, some but not all of the participants (the pretest vs. no-pretest groups) were asked to solve a challenging problem before consulting Google for needed information. The results revealed that this pretesting method significantly enhanced participants' subsequent recall of the searched content. These findings highlighted the potential of pretesting as a search-for-learning strategy to deepen learning and improve memory retention. 

Building upon these findings, the present study aims to extend the investigation into the efficacy of pretesting as a learning strategy by using GenAI as the information retrieval tool. Specifically, we seek to examine whether pretesting, when implemented with a GenAI model, such as ChatGPT, can yield similar enhancements in participants' subsequent recall of the searched content. By exploring the potential of ChatGPT as a learning aid, we aim to contribute to a deeper understanding of optimal learning strategies in the age of AI.

\section{Theoretical Background}
\label{sec:headings}
\subsection{Pretesting as a Catalyst for Learning}
Pretesting (also called pre-questioning) is the strategy of testing learners on new material before formal instruction, with the aim of activating prior knowledge and priming future learning. At first glance, quizzing students on topics they have not yet learned may seem counterintuitive; however, a growing body of research shows that such pretests can significantly benefit subsequent learning. This phenomenon, known as the pretesting effect, has been demonstrated across a variety of learning materials, such as texts \cite{richland2009pretesting, sana2023broader} and videos \cite{pan2020learning}, even though learners’ initial answers are often wrong. In essence, attempting to answer questions in advance prepares learners for upcoming content. It triggers psychological processes and exposes knowledge gaps, thereby heightening attention to the to-be-learned information when it is later studied \cite{pan2023prequestioning}. Recent comprehensive reviews conclude that pretesting frequently enhances both memory retention and transfer of learning, though the magnitude of this benefit can depend on how the pretest is implemented and what outcomes are measured \cite{pan2023prequestioning}. This robust evidence establishes pretesting as a valuable pedagogical tool for boosting learning, and it invites deeper inquiry into how and why this strategy works – especially as new technologies open novel ways to apply it.

\subsection{Cognitive Mechanisms Underlying Pretesting}
The learning gains from pretesting are explained by multiple theoretical accounts, reflecting recent advances in our understanding of why pre-instruction testing potentiates learning. One account emphasizes the role of errorful generation - when learners attempt answers and generate errors, those incorrect responses can function as productive “missteps” that engage learners more actively in learning. For example, even an incorrect guess can serve as a semantic mediator linking the question and the correct answer \cite{vaughn2012guessing, huelser2012making}; when feedback or the correct answer is later provided, the learner can connect it to their prior guess, strengthening memory for the right answer \cite{pan2021pretesting}. Another account, known as the search set theory, posits that a prequestion prompts learners to activate a set of related concepts from memory (a search set), which likely includes the correct information or closely associated knowledge \cite{grimaldi2012and}.When the correct answer is subsequently encountered, it is encoded more deeply because those memory networks have been pre-activated. These and other perspectives (e.g., heightened attention to knowledge gaps, and improved encoding of feedback) converge on the idea that pretesting changes how learners engage with subsequent study.

The attentional window hypothesis proposed by Sana and Carpenter \cite{sana2023broader} offers a compelling explanation that potentially unifies various theoretical accounts of pretesting effects. According to this hypothesis, pretesting opens a temporary attentional window during which learners actively search for answers to pretest questions until those answers are encountered, at which point the window closes. This temporal perspective on attention allocation helps explain why pretesting benefits vary across different learning situations and materials. The hypothesis suggests that pretesting enhances processing of information encountered while this attentional window is open, with the scope of benefits determined by when relevant information appears in the learning materials. This attentional mechanism likely operates through increased interest, curiosity, and deeper cognitive engagement with material encountered during the active search period created by pretesting.

A comprehensive three-stage framework has been proposed to integrate various theoretical perspectives on pretesting effects \cite{pan2023prequestioning}. This framework conceptualizes how pretesting enhances learning through distinct cognitive processes occurring at different points in the learning sequence. In the initial stage, pretesting activates unique psychological processes not triggered by conventional study methods. During the second stage, when learners encounter the actual learning material, these activated processes optimize information encoding through enhanced attention and strategic information seeking. Finally, at assessment, improved performance emerges from this enhanced encoding and potentially from the retrieval pathways established during pretesting. This framework suggests pretesting may influence learning through multiple routes simultaneously: by indirectly enhancing overall engagement with learning materials, by directly triggering specific information-seeking behaviors, or by creating retrieval cues that support later recall even without significantly altering learning behaviors. This integrated perspective acknowledges that multiple cognitive mechanisms likely contribute to pretesting effects, with their relative importance varying based on learning materials and implementation approaches.

\subsection{Pretesting in the Digital Information Age}
In today’s computer-enabled learning environments, students have unprecedented on-demand access to external information resources. The Internet, in particular, functions as a kind of “transactive memory” partner. In this partnership, the Internet holds all the knowledge while the human partner leverages it for effortless external retrieval \cite{fisher2015searching}. Rather than remembering facts, human partners often remember where to find them (e.g., via a quick search), effectively outsourcing part of their memory to digital tools. While this availability of information is convenient, it raises concerns from a pedagogical perspective: this instant access to information may short-circuit the desirable cognitive efforts triggered by strategies (e.g., pretesting) that potentiate the learning of new information. 

If a student can immediately Google an answer, they may skip the productive struggle of trying to recall or reason it out – a process which, as noted above, can strengthen learning. Recent research on “pretesting before searching” suggests a way to adress these concerns. Empirical studies show that integrating a pretest step before using internet resources markedly improves learning outcomes. Giebl et al.  \cite{giebl2021answer}, for example, found that learners who attempted to answer questions prior to web search later recalled both the sought information and related content significantly better than those who searched for answers straightaway. This benefit was especially pronounced for learners who had some prior knowledge of the topic, suggesting that pretesting helps activate existing knowledge, which in turn allows new information from the web to be integrated more meaningfully. Extending this line of work, Giebl et al. 
\cite{giebl2023thinking} demonstrated that prompting students to “think before Googling” (i.e., answer from memory first) leads to higher subsequent recall and understanding than an immediate lookup. In essence, even in the age of ubiquitous search engines, initial thought holds greater value than instant answer. Pretesting fits perfectly into this paradigm by ensuring that learners engage actively with a question on their own before tapping into external databases or tools. Thus, far from being made obsolete by Google, pretesting may be even more important in digital learning, serving as a cognitive scaffold that keeps students mentally engaged when using computers for information retrieval.

\subsection{Generative AI: New Opportunities and Challenges}
Generative AI tools like ChatGPT represent the latest leap in computer-assisted learning. These AI systems go beyond static search results: they can hold open-ended conversations, tailor their responses to a learner’s needs, and provide instant, context-specific explanations. Educationally, this is a double-edged sword. On one hand, generative AI offers an unprecedented level of interactivity and personalization in learning. A tool such as ChatGPT can act as a virtual tutor available 24/7, capable of answering questions, rephrasing explanations, and guiding students through problems in a dialogic manner. Such capabilities open up exciting pedagogical possibilities. For instance, an AI could adapt pretest questions on the fly based on a student’s prior responses, offering hints or increasing challenge as needed. On the other hand, the convenience and power of AI tools might tempt students to bypass productive thinking processes. If a detailed answer or solution is just a prompt away, learners may become overly reliant on the AI, potentially undermining the benefits of struggle and self-explanation that educators know are vital for learning (e.g.,\cite {sinha2021problem}). Early investigations into ChatGPT’s educational impact reflect this tension. A recent systematic review found that ChatGPT-based interventions improve students’ performance and even higher-order thinking propensity, but at the same time reduce learners’ mental effort – with no significant gain in self-efficacy – compared to traditional learning experiences \cite{deng2024does}. These developments underscore a pressing question for computer-assisted education: How can we leverage AI’s strengths (e.g., rich information, adaptive feedback) without negating the cognitive processes that build durable understanding? This study posits that pretesting is a prime candidate to meet that challenge.

\section{Present Study}
Given the proven benefits of pretesting and the rise of generative AI in education, a timely opportunity emerges to combine the two. We argue that incorporating pretesting into AI-supported learning can preserve the productive struggle that drive learning, and then use the AI to provide feedback, clarification, or extended guidance. This approach is pedagogically valuable because it ensures that the computer is used not just as a cognitive partner storing needed information, but as a tool for amplifying effective learning strategies. The present study situates itself at the intersection of these developments.

Building on recent findings that pretesting enhances learning even when students use online resources, we extend the investigation to generative AI. Across three interconnected studies, we explored how pretesting enhances learning outcomes and whether the adaptability of AI-assisted pretesting, task characteristics, and student motivation influence its efficacy.

In Study 1, we examined whether pretesting improves learning outcomes when students use a GenAI tool, such as ChatGPT, to solve problems. This study aimed to determine if pretesting helps activate relevant prior knowledge, thereby facilitating better engagement with and retention of information provided by the GenAI. Participants were assigned to either a pretesting group or a no-pretesting group and completed a task that required executing a linear, well-structured sequence of calculations with ChatGPT to systematically analyze the given data. This initial investigation provided a foundation for understanding how pretesting interacts with generative AI tools to support learning.

In Study 2, we extended this investigation by exploring whether AI-assisted pretesting—where the AI dynamically adapts pretesting questions based on student responses—provides any advantage over traditional pretesting, which uses a fixed set of pre-written questions and does not offer follow-up clarification or additional inquiries. The task from Study 1 was also utilized in Study 2.This investigation was motivated by the premise that AI-assisted pretesting offers a more tailored and interactive learning experience, which could better engage students and address misinterpretations in real-time. 

Recognizing that task characteristics might influence the effectiveness of pretesting strategies, we designed Study 3 to investigate whether the effectiveness of AI-assisted pretesting depends on the type of task used and the role of student motivation. Specifically, we shifted from linear, well-structured analytical tasks to iterative, exploratory evaluation tasks involving model selection, where students evaluate multiple scenarios and make reasoned decisions based on various criteria. By considering both task type and motivational factors, this study extended the scope of our inquiry to understand the broader applicability and potential limitations of AI-assisted pretesting. To provide a clear overview of the evolution of our investigation and the key differences between the three studies, we summarized their focus, task designs, group assignments, and contributions in Table 1.

\begin{table}[htbp]
\centering
\vspace{-20pt}
\caption{Study Comparison Table}
\renewcommand{\arraystretch}{1.2}
\begin{tabularx}{\textwidth}{|>{\raggedright\arraybackslash}X|>{\raggedright\arraybackslash}X|>{\raggedright\arraybackslash}X|>{\raggedright\arraybackslash}X|}
\hline
\textbf{Dimension} & \textbf{Study 1} & \textbf{Study 2} & \textbf{Study 3} \\
\hline
\textbf{Focus} & 
Investigate the effect of pretesting with ChatGPT. & 
Compare static (fixed) pretesting to AI-assisted pretesting. & 
Examine the effectiveness of static vs. AI-assisted pretesting in an iterative, exploratory task. \\
\hline
\textbf{Pretesting Design} & 
Traditional, static pretesting. & 
Static (fixed) vs. AI-assisted (dynamic) pretesting. & 
Static (fixed) vs. AI-assisted (dynamic) pretesting. \\
\hline
\textbf{Task Type} & 
Linear, well-structured analytical task. & 
Same task as Study 1. & 
Iterative, exploratory evaluation task requiring reasoning and decision-making. \\
\hline
\textbf{Group Assignment} & 
Pretest vs. no-pretest. & 
Fixed pretest group vs. AI-assisted pretest group. & 
Fixed pretest group vs. AI-assisted pretest group. \\
\hline
\textbf{Research Goal} & 
Establishes the baseline for AI-assisted pretesting. & 
Evaluates the benefits of AI-assisted pretesting. & 
Explores task and motivation interactions with AI-assisted pretesting. \\
\hline
\end{tabularx}
\vspace{0pt}
\end{table}

Taken together, these three studies present a comprehensive investigation into the effectiveness of pretesting as a learning strategy in AI-supported education. They build upon one another to systematically examine critical factors—including the adaptability of AI-assisted pretesting, the nature of the task, and student motivation—that contribute to its success. By presenting these studies in a single manuscript, we aim to provide an integrated perspective on how pretesting strategies can be optimized to support learning in GenAI ecosystems.

\section{Study 1}
\subsection{Method}
\subsubsection{Experimental Design}
The study, designed as a true-experimental posttest with randomized groups \cite{field2002design}, was carried out in an intermediate-level statistics course and consisted of two phases. In the first phase, all participants engaged with materials covering one-way chi-square analysis. This was followed by a second phase, where participants faced a complex cybersecurity problem that required them to conduct a two-way chi-square analysis. The problem introduced in the second phase was deliberately designed to be unsolvable using only the information from the first phase. This decision was informed by the teaching team’s extensive experience in instructing this course. This approach ensured that while students had a basic understanding of chi-square analysis, their knowledge was insufficient for applying two-way chi-square analysis to more complex problems. This setup allowed students to engage with more complex issues based on their preliminary knowledge, facilitating the establishment of a pretesting task within the study design.

Specifically, participants in the no-pretest group were immediately given access to the necessary information through ChatGPT. Conversely, those in the pretest group attempted to solve the problem independently before they could access this additional help. The study concluded with a  multiple-choice test to evaluate not only the retention and comprehension of the learned content but also the participants' ability to apply this knowledge in new situations.

\subsubsection{Participants}
Seventy-three undergraduate students majoring in information sciences at a large public research university in Pennsylvania, United States, were initially recruited for the study. The cohort comprised 17 women and 56 men. Technical issues, such as internet connectivity and user account problems, along with non-adherence to the instructions, led to the exclusion of 12 participants. Consequently, 61 students completed the study, with 31 (50.8\%) assigned to the no-pretest group and 30 (49.2\%) to the pretest group. The study received approval from the Institutional Review Board (IRB No 00024664).

\subsubsection{Procedure}
The experiment comprised two phases, as illustrated in Figure 1. Both phases occurred within the same week. 
The course, which convened twice weekly, allotted 75 minutes per session. Phase 1 occurred during the initial session, focusing on one-way chi-square analysis. Phase 2 unfolded in the subsequent session, starting with a recall test to measure participants' understanding of the concepts covered in Phase 1. Subsequently, participants completed a more complex task requiring the use of two-way chi-square analysis, for which they had not yet received all the necessary information.
During Phase 2, participants were randomly assigned to either a pretest group or a no-pretest group. Those in the pretest group attempted to complete the task before accessing additional information through ChatGPT, while those in the no-pretest group could access the information immediately using ChatGPT. This design aimed to evaluate whether attempting the task before engaging with ChatGPT for needed information (pretest) impacted learning compared to prompting ChatGPT immediately without attempting the task (no pretest).To assess the effects of pretesting on learning in the context of GenAI-supported environments, a final multiple-choice test containing questions related to a different but conceptually similar problem scenario was administered to both groups. This test was designed to evaluate participants' understanding of concepts presented in Phase 2.

\begin{figure}[h!]
  \centering
  \includegraphics[width=0.75\textwidth]{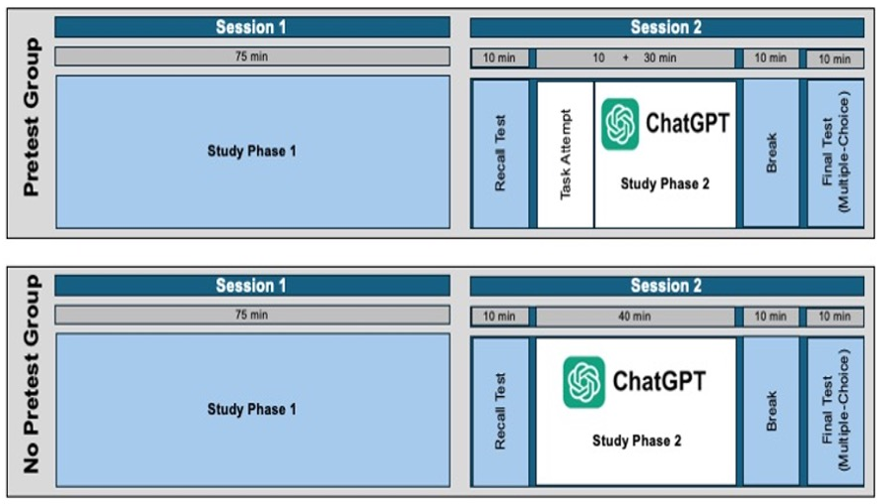} % Specify the image file here
 \caption{Study 1 Procedure} % Caption for the figure
 \label{fig:figure1} % Label for referencing in text

\end{figure}

\subsubsection{Materials}
\paragraph{Phase 1 Activity.} During Phase 1, participants received instructions on the fundamental aspects of chi-square tests, including (a) the definition of a chi-square test, (b) the rationale behind using chi-square tests, (c) the calculation and interpretation of chi-square values, and (d) the application of one-way chi-square tests to cybersecurity problems. Following the instructions, participants collaborated in groups to apply their knowledge by solving a practical problem.% (see Figure 2).

%\begin{figure}[h!]
 %   \centering
    %\vspace{-20pt}\includegraphics[width=0.75\textwidth]{Figure2.png} % Specify the image file here
    %\caption{Scenario used in Phase 1} % Caption for the figure
    %\label{fig:figure2} % Label for referencing in text
    %\vspace{-20pt}
%\end{figure}

\paragraph{Recall Test.}The recall test aimed to assess students' understanding of key concepts related to chi-square tests, as taught in Phase 1. This test was administered to both the No-Pretest Group and the Pretest Group. Participants were presented with a scenario and asked to answer multiple-choice questions based on the scenario. The outcomes of this recall test were employed to assess for any statistically significant differences between the No-Pretest Group and the Pretest Group. 

%\begin{figure}[h!]
 %   \centering
    %\includegraphics[width=0.75\textwidth]{Figure3.png} % Specify the image file here
    %\caption{Scenario and two of the items used in the recall test.} % Caption for the figure
    %\label{fig:figure3} % Label for referencing in text
%\end{figure}

\paragraph{Phase 2 Activity.} During Phase 2, participants were presented with a cybersecurity problem scenario that required the use of two-way chi-square analysis. This task built upon concepts studied in Phase 1, such as calculating chi-square values and finding critical values for assessing statistical significance. However, it also required information only available via ChatGPT, such as calculating expected frequencies and determining degrees of freedom for a two-way chi-square test.

Participants assigned to the pretest group did not have immediate access to ChatGPT. Instead, they were instructed to attempt to solve the task without assistance for a period before being given access to ChatGPT. No corrective information was provided during their problem-solving attempts. To ensure that participants engaged cognitively with the problem, participants were asked to answer questions in the pretest (see Figure 2 for pretest questions). Participants who did not provide answers to these questions were excluded from the study. The information necessary to solve the task could be accessed via ChatGPT, which became available to participants in the pretest group once they had submitted their answers to the questions in the pretest.

\begin{figure}
    \centering
    \includegraphics[width=\textwidth]{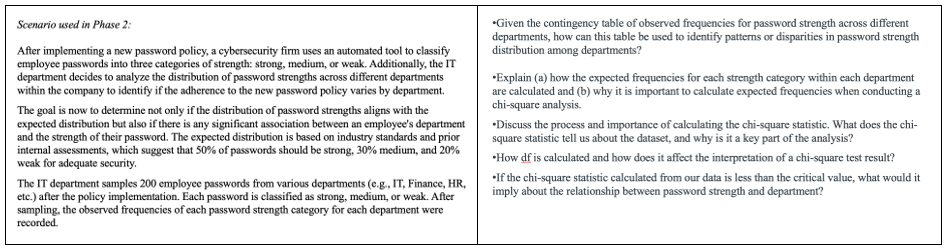}
    \caption{Scenario used in Phase 2 and pretest questions}
    \label{fig:figure4}
   
\end{figure}

In contrast, participants in the no-pretest group could access ChatGPT immediately to find the relevant information necessary to solve the task. To ensure consistency in the information accessed and to control for potential variations in prompting strategies, participants were provided with sample prompts and comprehensive questions. 
%(see Figure 4). 
These prompts were designed to guide participants through the analytical process step-by-step, ensuring that they engaged with the information in a systematic manner.

Participants were instructed to follow a specific sequence when interacting with ChatGPT:
\begin{enumerate}
    \item \textbf{Introduce the Data:} Begin by clearly communicating the dataset and the problem to ChatGPT using the preliminary step prompt

    \item \textbf{Step-by-Step Engagement:} Progress through each analytical step using the provided direct prompts. This approach will help you methodically understand and apply the chi-square analysis to the dataset.

    \item \textbf{Comprehension and Reflection:} Answer the comprehension questions following each step. These should be in your own words and reflect a deep understanding of the statistical process and its implications.

    \item \textbf{Concluding Reflections:} After completing the analysis, reflect on the significance of your findings and consider their implications in the context of cybersecurity and beyond.
\end{enumerate}

%\begin{figure}[h!]
 %   \centering
%\vspace{-20pt}    \includegraphics[width=0.75\textwidth]{Figure5.png} % Specify the image file here
 %   \caption{Preliminary step and first two steps with prompts and comprehensive questions used in Phase 2 activity.} % Caption for the figure
  %  \label{fig:figure5} % Label for referencing in text
   % \vspace{-20pt}
%\end{figure}

To prevent direct copying and pasting from ChatGPT in their responses, participants were required to include a copy of their conversation with ChatGPT in their submissions. Once the conversation histories were collected, they were examined to ensure that participants responded to comprehension questions by summarizing the ChatGPT interaction in their own words rather than directly copying and pasting from the conversation.

\paragraph{Final Test.} The final test was administered to both the no-pretest group and the pretest group. Participants were presented with a scenario and asked to answer multiple-choice questions based on the scenario. The final test materials consisted of 11 multiple-choice transfer questions on concepts from both study phases.

%\begin{figure}[h!]
    %\centering
    %\includegraphics[width=0.75\textwidth]{Figure6.png} % Specify the image file here
    %\caption{Scenario and two of the items used in the final test.} % Caption for the figure
    %\label{fig:figure6} % Label for referencing in text
%\end{figure}

\subsubsection{Data Analysis}
The normal distribution of recall test and final test scores were examined. The final test results showed skewness (-.059) and kurtosis (-.548) values of $\pm$1.5, indicating no issue \cite{tabachnick2013using}. The recall test results were -1.593 and 2.296, respectively, with one outlier detected in the experimental group. After it was removed, the skewness and kurtosis values were reduced to -1.295 and.657, respectively, which fell below the threshold. As a result, data from 31 participants in the no-pretest group and 29 participants in the pretest group were used in the analysis. 

\paragraph{Recall Test Performance} We assessed the recall test performance using an independent t-test, treating the group (pretest group or no-pretest group) as the categorical independent variable and the recall test scores as the continuous dependent variable. The results indicated no significant difference in average scores between the two groups, t(58) = 0.584, p = 0.562). As indicated in Figure 3a, the descriptive statistics for the pretest group were M = 71.03, SD = 13.72, and for the no-pretest group, M = 72.90, SD = 11.01. These findings suggest that participants had a similar level of understanding of the one-way chi-square analysis concepts covered in Phase 1, indicating that this prior knowledge was unlikely to significantly impact the results of the subsequent test related to two-way chi-square analysis.

\begin{figure}[h!]
    \centering
  
    \includegraphics[width=\textwidth]{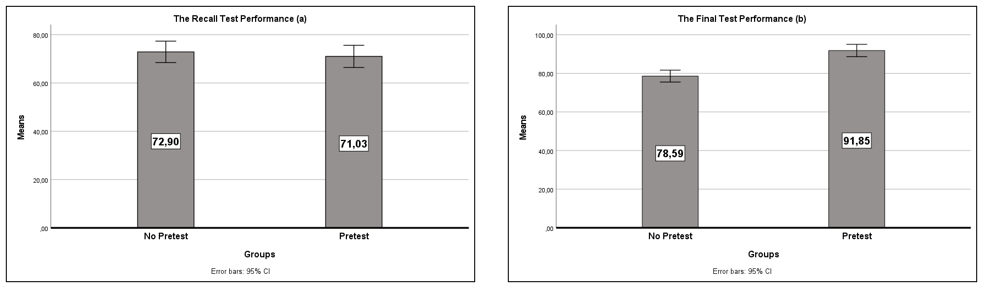} % Specify the image file here
    \caption{The Recall and Final Test Performances for No-pretest and Pretest groups} % Caption for the figure
    \label{fig:figure7} % Label for referencing in text

\end{figure}

\paragraph{Final Test Score Analysis} Based on the similarity in recall test performance between the two groups, we proceeded with an independent t-test. In this test, the students' final test scores were treated as the continuous dependent variable, while the group they were assigned to (pretest group or no-pretest group) was considered the categorical independent variable.

\subsection{Results}
The independent sample t-test results revealed a significant difference between the pretest and no-pretest groups in the final test scores, t(58) = 5.979, p < .001, Cohen’s d = 1.543, indicating a large effect size \cite{vacha2004estimate}. As shown in Figure 3b, the mean score for the no-pretest group was 78.59 (SD = 7.62), while that of the pretest group was 91.85 (SD = 9.52). 

\subsection{Discussion}
The findings from this study suggest that pretesting before engaging with conversational AI tools can significantly enhance learning outcomes. Specifically, students who engaged in pretesting demonstrated a better ability to solve problems and apply statistical concepts compared to those who directly accessed AI tools without pretesting. These results align with prior research showing that pretesting activates prior knowledge and prepares learners to engage more deeply with subsequent information \cite{carpenter2011semantic, grimaldi2012and, pan2023prequestioning}.

One possible explanation for the observed benefits of pretesting lies in its ability to prime students' cognitive processes. By attempting to solve problems independently before accessing AI support, students are likely to activate relevant knowledge schemas, which facilitates deeper engagement during subsequent problem-solving with AI tools. This aligns with the theory of retrieval-based learning \cite{karpicke2014retrieval}, which posits that attempting to retrieve information strengthens memory traces and enhances learning efficiency.

These results underscore the potential for merging educational strategies like pretesting with modern generative AI. This integration requires further exploration to determine how AI can best complement pretesting across various disciplines.

%This study also highlights the potential for integrating traditional educational strategies, like pretesting, with modern generative AI tools. Conversational AI, such as ChatGPT, may amplify the benefits of pretesting by providing learners with tailored questions and interactive learning experiences. However, this integration requires further investigation to explore the extent to which AI’s capabilities can complement pretesting across various disciplines and tasks.

\section{Study 2}
\subsection{Experimental Design}
This study employed a true-experimental posttest design with randomized groups to compare the effectiveness of two pretesting strategies in an intermediate-level statistics course. While the overall structure and phases of the study mirrored the design of Study 1, the grouping variable was adjusted to evaluate the impact of fixed versus AI-assisted pretesting approaches.

Participants were randomly assigned to one of two groups:

\begin{enumerate}
    \item \textbf{Fixed Pretesting (FP) group:} Participants in this group attempted to answer a fixed set of 4-5 pre-written questions related to the complex cybersecurity problem independently. These questions were presented in a predetermined sequence, and participants worked through them without any adaptation based on their responses.
    \item \textbf{AI-Assisted Pretesting (AP) group:} Participants in this group interacted with an AI Coach that dynamically generated questions tailored to their responses. Rather than following a fixed order, the questions evolved based on student input, creating an interactive dialogue aimed at addressing specific misinterpretations or gaps in knowledge in real-time.
\end{enumerate}
The AI-assisted pretesting approach was designed to provide a personalized learning experience, leveraging the AI Coach's ability to adjust the difficulty, content, and focus of questions dynamically. This interactive dialogue emphasized flexibility and targeted support compared to the static nature of fixed pretesting.

As in Study 1, the experiment concluded with a multiple-choice test to assess retention, comprehension, and the ability to transfer knowledge to novel contexts. This design allowed for a robust comparison of fixed versus AI-assisted pretesting strategies within a controlled environment, shedding light on the potential benefits of dynamic question generation and interactive dialogue in pretesting tasks.

\subsection{Participants}
The study was conducted in the same institution and course as Study 1, but during the subsequent semester. A total of 78 undergraduate students completed the study after accounting for exclusions due to technical issues and non-adherence to instructions. The Fixed Pretesting (FP) group consisted of 36 students (46.2\%), while the AI-Assisted Pretesting (AP) group included 42 students (53.8\%). The cohort comprised 16 female students (20.5\%) and 62 male students (79.5\%).

\subsection{Procedure and Materials}
The procedure and materials for Study 2 followed the same overall structure as Study 1, consisting of two phases conducted within the same week during a course with two 75-minute sessions. Adjustments were made to accommodate the grouping variable introduced in the experimental design section, focusing on fixed versus AI-assisted pretesting strategies.

\textbf{Phase 1.} During the first session, all participants engaged with instructional materials on one-way chi-square analysis to establish foundational knowledge. These materials were identical to those used in Study 1, ensuring consistency across studies.

\textbf{Phase 2.} The second session began with a recall test designed to assess participants' understanding of the concepts from Phase 1. The recall test was the same as in Study 1. After completing the recall test, participants engaged in a pretesting activity where they worked on a complex cybersecurity problem requiring a two-way chi-square analysis. The pretesting component differed between the two groups: one group completed a fixed set of pre-written questions identical to those used in Study 1 (Fixed Pretesting, FP), while the other engaged in an interactive dialogue with an AI Coach that dynamically adapted questions to their responses (AI-Asssisted Pretesting, AP). To ensure the AI Coach’s adaptability, we requested that students share their ChatGPT conversation history; we then reviewed these conversations to confirm that the system was tailoring its prompts to individual responses. After completing the pretesting activities, all participants accessed ChatGPT to gather information necessary for completing the task presented in the pretesting activity. This setup allowed for a direct comparison of the two pretesting strategies under consistent conditions. All participants completed a final multiple-choice test at the end of Phase 2. This test assessed learning outcomes and was identical to the final test employed in Study 1. (see Figure 4 for Study 2 procedure). 

%The procedure for Study 2 followed the same overall structure as Study 1, consisting of two phases conducted within the same week during a course with two 75-minute sessions. Phase 1 and Phase 2 took place during the first and second sessions, respectively, with adjustments reflecting the grouping variable in Study 2.

%In Phase 1, all participants studied materials on one-way chi-square analysis to establish foundational knowledge.

%Phase 2 began with a recall test assessing participants' understanding of the concepts from Phase 1. Following this, participants were randomly assigned to one of two groups. Depending on their assigned group, participants either attempted a fixed set of pre-written questions or engaged with dynamically generated questions through an AI Coach before accessing ChatGPT to obtain the information needed for a complex cybersecurity task requiring two-way chi-square analysis. This task was designed to test their ability to apply and extend their knowledge in a novel context.

%To evaluate learning outcomes, a final multiple-choice test (i.e., exit test) was administered to all participants. The test included questions on a new but conceptually similar problem scenario to measure the retention and transfer of knowledge gained in Phase 2 (see Figure 8 ).

\begin{figure}[h!]
    \centering

    \includegraphics[width=\textwidth]{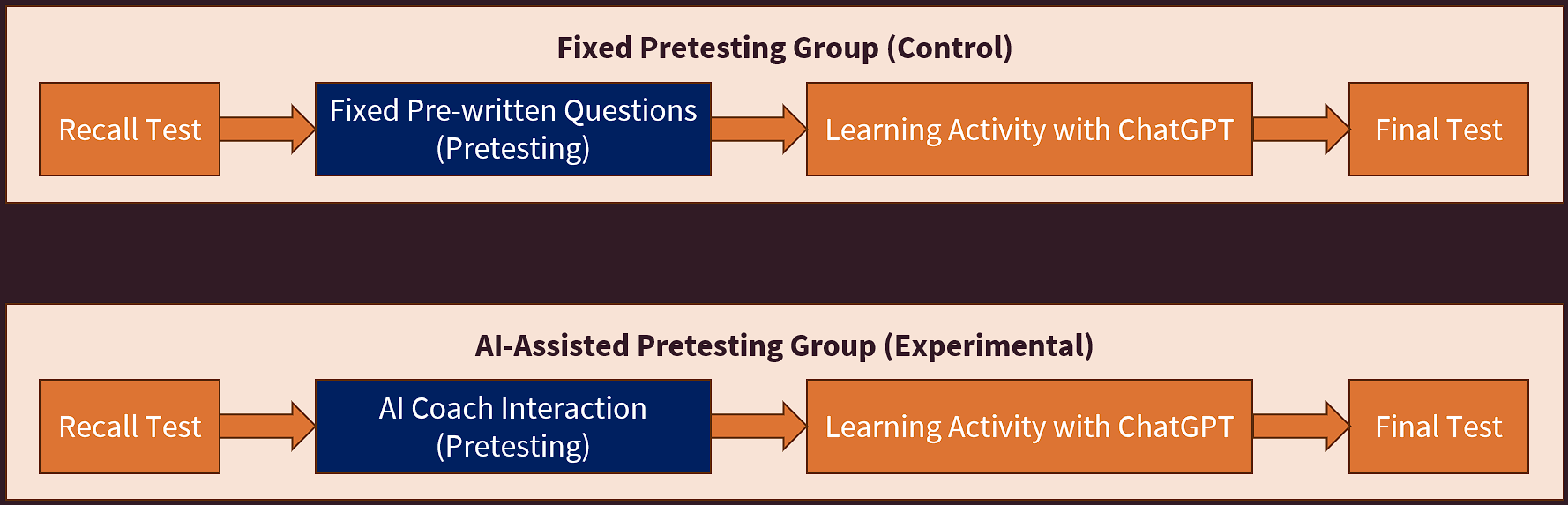} % Specify the image file here
    \caption{Study 2 Procedure - Phase 2} % Caption for the figure
    \label{fig:figure8} % Label for referencing in text

\end{figure}

This design preserved consistency with Study 1 while introducing the adaptive and fixed pretesting strategies for comparison.

\subsection{Analysis}
The normality of recall test and final test scores was examined prior to analysis. Initial inspection revealed skewness and kurtosis values exceeding the acceptable threshold of ±1.5, indicating potential distribution issue \cite{tabachnick2013using}. Specifically, the recall test had skewness (-1.230) and kurtosis (3.573), while the final test showed skewness (-2.060) and kurtosis (5.355). Outlier analysis identified two outliers in the AI-Assisted Pretesting (AP) group. These outliers were removed, reducing skewness and kurtosis values to within acceptable ranges. The adjusted skewness values were -0.191 for the recall test and -0.098 for the final test, while kurtosis values were -1.437 and 1.353, respectively.

Following this adjustment, data from 42 participants in the Fixed Pretesting group and 36 participants in the AI-Assisted Pretesting group were included in the analysis.

\paragraph{Recall Test Performance} To evaluate recall test performance, an independent t-test was conducted, with group assignment (Fixed Pretesting and AI-Assisted Pretesting groups) as the categorical independent variable and recall test scores as the continuous dependent variable. The results revealed no significant difference in average scores between the two groups, t(76)=0.385, p=0.701. Descriptive statistics indicated that the Fixed Pretesting group had a mean score of M = 70.48 (SD=13.43), while the AI-Assisted Pretesting group scored M=71.67 (SD=13.84), as shown in Figure 5a.

These findings suggest that participants in both groups had a comparable level of understanding of the one-way chi-square analysis concepts covered in Phase 1. Consequently, prior knowledge is unlikely to have significantly influenced performance on the subsequent tasks related to two-way chi-square analysis.

\begin{figure}[h!]
    \centering
    \includegraphics[width=\textwidth]{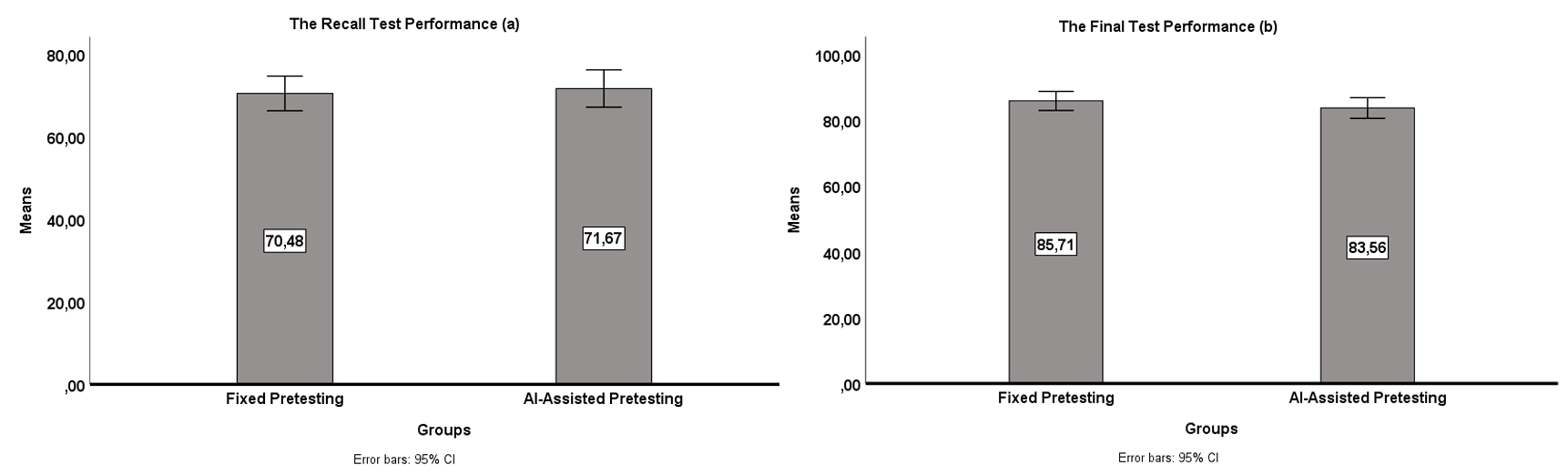} % Specify the image file here
    \caption{The Recall and Final Test Performances for Fixed Pretesting and AI-Assisted Pretesting groups (Study 2)} % Caption for the figure
    \label{fig:figure19} % Label for referencing in text
   
\end{figure}

\paragraph{Final Test Score Analysis} 
Given the similarity in recall test performance between the two groups, an independent t-test was conducted. In this analysis, the students' final test scores were treated as the continuous dependent variable, while the assigned group (Fixed Pretesting or AI-Assisted Pretesting) served as the categorical independent variable.

\subsection{Results}

The independent t-test results indicated no significant difference in performance between the two groups, t(76)=1.002,p=0.319. As illustrated in Figure 5b, the Fixed Pretesting group had a mean score of M=85.71 (SD=8.49), while the AI-Assisted Pretesting group scored M=83.56 (SD=10.44).

\subsection{Discussion}
The findings from Study 2 revealed no significant differences in final test performance between the Fixed Pretesting (FP) and AI-Assisted Pretesting (AP) groups. These results suggest that both pretesting strategies—whether involving a fixed set of pre-written questions or dynamic, adaptive interaction with an AI Coach—were similarly effective in preparing students to tackle the complex problem-solving tasks assessed in the final test.

One potential explanation lies in the computationally focused nature of the Phase 2 pretesting task, which emphasized chi-square calculations (e.g., expected vs. observed frequencies). This rigid, formulaic structure may inherently limit the advantages of adaptive pretesting. While the AI Coach adapted questions, students’ ability to fully benefit from this adaptability might have been constrained by the task’s well-structured, formula-driven focus. In contrast, fixed pretesting may have been adequate to prepare learners for the computational demands of the task.
%One potential explanation lies in the nature of the task used in Phase 2, which formed the basis of the pretesting activity. The task required students to engage with the chi-square formula by focusing on its components, such as expected and observed frequencies, and performing calculations to solve problems. This computationally focused task emphasized the ability to compute values and interpret formula components. Such tasks may inherently provide less opportunity for adaptive pretesting to demonstrate its advantages. Although the AI Coach dynamically tailored questions to individual responses, the rigid structure and formulaic nature of the task might not have allowed students to fully benefit from this adaptability. Fixed pretesting, which also engaged students with the formula’s parameters, may have been sufficient to prepare students for the computational demands of the task.

This raises an important question: Does the nature of the task—linear and well-structured versus iterative and exploratory—moderate the effectiveness of pretesting strategies in GenAI-supported learning? Tasks requiring higher-order thinking (e.g., evaluating options, integrating multiple factors) may better reveal the potential advantages of AI-assisted pretesting.
%does the nature of the task — such as a linear, well-structured analytical task versus an iterative, exploratory task — moderate the effectiveness of pretesting strategies, particularly in the context of GenAI-supported learning environments? Tasks that require higher-order thinking, such as evaluating options, making decisions, and integrating multiple factors, may provide a more sensitive test of the potential advantages of AI-assisted pretesting.

The findings of Study 2 thus highlight the need to examine how task characteristics interact with pretesting approaches, setting the stage for a more nuanced exploration in Study 3.

\section{Study 3}
\subsection{Experimental Design}
Study 3 followed the same experimental design as Study 2, utilizing a true-experimental posttest design with randomized groups to investigate the effectiveness of pretesting strategies in an intermediate-level statistics course. The study aimed to build on the findings of Study 2 by examining whether the nature of the pretesting task influenced the effectiveness of fixed versus AI-assisted pretesting approaches.

The primary distinction in Study 3 was the iterative, exploratory nature of the pretesting task. Unlike the linear, well-structured analytical task used in Study 2, this task required participants to engage in model selection by evaluating multiple scenarios and making decisions based on various criteria. This shift added complexity to the learning process, requiring students to balance multiple considerations while interacting with generative AI. While simpler tasks may not induce strong feelings of anxiety or demand much autonomy, this new task placed students in a decision-intensive setting where they had to make judgments with AI support. 
%This task design provided an opportunity to assess whether AI-assisted pretesting offered advantages in tasks requiring deeper conceptual engagement and iterative reasoning.

To better understand how students navigated this exploratory learning process, we incorporated the Intrinsic Motivation Inventory (IMI), originally developed by McAuley et al. \cite{mcauley1989psychometric} and later adapted for AI-based learning contexts by Yin et al. \cite{yin2021conversation}. This questionnaire assessed five key motivational dimensions: interest-enjoyment, tension-pressure, perceived choice, perceived competence, and perceived value. By capturing these aspects, we aimed to examine not only students' learning gains but also their engagement, sense of autonomy, confidence in their abilities, and overall perception of the learning experience within the AI-assisted pretesting environment.

%While simpler tasks may not spark strong feelings of anxiety or demand much autonomy, the new environment presented students with multiple scenarios and required them to make informed decisions with the support of generative AI. This added complexity led us to utilize the specific dimensions adapted from the Intrinsic Motivation Inventory, originally developed by McAuley and colleagues (1989) and later refined for AI-based learning contexts by Yin, Goh, Yang, and Xiaobin (2021). By including scales for interest-enjoyment, tension-pressure, perceived choice, perceived competence, and perceived value, we could capture how students navigated this exploratory, decision-intensive process. For instance, interest-enjoyment uncovered whether they found these iterative tasks engaging or tedious, tension-pressure illuminated if they felt calm or stressed when evaluating multiple models, perceived choice addressed the level of autonomy they experienced, perceived competence gauged their confidence in selecting and justifying certain models, and perceived value revealed the extent to which they deemed the entire learning method beneficial and worthwhile. These focused measures offered deeper insight into the practical and emotional aspects of student motivation, illuminating how changes in both task structure and the integration of GenAI tools can shape learners’ engagement, confidence, and overall sense of purpose.

As in Study 2, participants were randomly assigned to one of two groups: the Fixed Pretesting (FP) group or the AI-Assisted Pretesting (AP) group. In the FP group, participants answered a fixed set of pre-written questions, while in the AP group, participants engaged in a dynamic, conversational session with an AI Coach that adapted questions based on their responses. Both pretesting strategies encouraged reflection without providing direct explanations. The experiment concluded with a final multiple-choice test to assess learning outcomes.

\subsection{Participants}
Study 3 was conducted in the same institution, course, and semester as Study 2. After accounting for exclusions due to technical issues and non-adherence to instructions, a total of 87 undergraduate students completed the study. Participants were randomly assigned to one of two groups: the Fixed Pretesting (FP) group (n=38, 43.7\%) and the AI-Assisted Pretesting (AP) group (n=49, 56.3\%). The cohort consisted of 21 women and 66 men.

\subsection{Procedure}
The procedure for Study 3 followed the same overall structure as Study 2, with  modifications to include both a procedural focus and the assessment of motivation. The key difference was the task introduced in Phase 2, which required participants to engage in model selection and related conceptual activities.

In \textbf{Phase 1}, all participants completed a recall test to assess their understanding of foundational concepts from prior coursework. This provided a baseline measure of prior knowledge to ensure comparability across groups.

In \textbf{Phase 2}, participants were randomly assigned to one of two groups explained in the experimental design section. After completing the pretesting activities, all participants engaged in the model selection task using ChatGPT. Following the ChatGPT activity, participants completed the final test to assess their ability to apply regression concepts to novel situations. Immediately afterward, participants completed the Learning Motivation Scale to provide data on their motivation during the study, including their engagement, interest, and perceived value of the pretesting activities.

\subsection{Materials}
The materials used in Study 3 were designed to evaluate participants’ understanding and application of regression analysis concepts while considering motivational factors, which were assessed as part of the study. These materials maintained consistency with the structure used in previous studies while incorporating regression-specific and motivational components.

\begin{enumerate}
    \item \textbf{Recall Test:} 
    A recall test was developed specifically for Study 3 to measure participants’ baseline knowledge of regression analysis. This test included questions about interpreting regression outputs, evaluating predictor significance, and understanding concepts like R² and adjusted R². The content differed from the recall test in Study 2, which focused on chi-square analysis, to align with the regression-specific focus of Study 3.
    
    \item \textbf{Pretesting Activity:} Participants in the Fixed Pretesting group answered a pre-written set of questions related to regression tasks, whereas participants in the AI-Assisted Pretesting group interacted with an AI Coach that dynamically tailored questions based on their responses (Figure 6) 
      \item \textbf{Regression Analysis Activity:} This task followed the same structured framework as in Studies 1 and 2, adapted for regression analysis. Participants engaged with ChatGPT to apply multiple regression to a practical cybersecurity scenario.
      \item \textbf{Final Test:} The final test, administered immediately after the regression analysis activity, assessed participants’ ability to transfer their knowledge of regression analysis to novel scenarios. The test included questions on model selection, multicollinearity, predictor evaluation, and model simplification.
      \item \textbf{Motivation Questionnaire:} Following the final test, participants completed a Motivation Questionnaire to assess their motivation during the study. The motivation questionnaire used in this study was adapted by Yin et al. \cite{yin2021conversation} from the Intrinsic Motivation Inventory (IMI) developed by McAuley et al.   \cite{mcauley1989psychometric}. It was designed to evaluate participants' subjective experiences of intrinsic motivation within the study's learning environments. The questionnaire comprised five scales: interest-enjoyment, tension-pressure, perceived choice, perceived competence, and perceived value. Participants responded to the questionnaire using a seven-point Likert scale, with options ranging from 1 (strongly disagree) to 7 (strongly agree).
\end{enumerate}

\begin{figure}[h!]
    \centering

    \includegraphics[width=\textwidth]{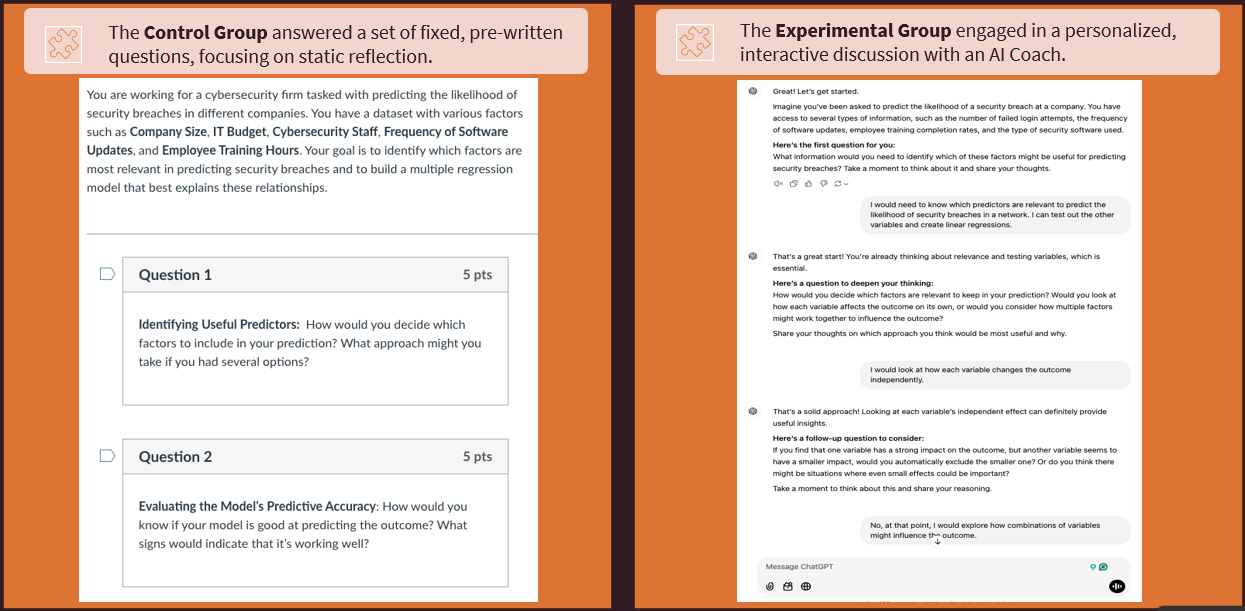} % Specify the image file here
    \caption{Pretesting Activity - Study 3} % Caption for the figure
    \label{fig:figure10} % Label for referencing in text

\end{figure}

%\begin{figure}[h!]
   % \centering
%\vspace{-10pt}    \includegraphics[width=0.60\textwidth]{Figure11.png} % Specify the image file here
 %   \caption{Scenario used in the regression activity} % Caption for the figure
 %   \label{fig:figure11} % Label for referencing in text
%    \vspace{-20pt}
%\end{figure}

\subsection{Analysis}
The normality of the recall test and final test scores was examined prior to analysis. For the recall test, skewness (-0.455) and kurtosis (-0.906) values fell within the acceptable threshold of ±1.5, indicating no issues \cite{tabachnick2013using}. For the final test, skewness (-1.082) and kurtosis (1.075) values fell within the acceptable threshold of ±1.5, indicating no issues. 
%Two outliers were detected in both the Fixed Pretesting (FP) and AI-Assisted Pretesting (AP) groups and subsequently removed. After their removal, skewness and kurtosis values for the final test were reduced to -0.562 and 0.957, respectively, falling within acceptable limits.
Additionally, three missing values were identified in the recall test and excluded from the analysis.

To analyze \textbf{recall test performance}, an independent t-test was conducted with one categorical independent variable (group: FP or AP) and one continuous dependent variable (recall test score), following the guidelines outlined by Field \cite{field2017discovering}. Assumptions for the t-test were tested using Levene’s Test for Homogeneity of Variances to ensure the robustness of the analysis. The results of the t-test indicated no significant difference in average recall test scores between the two groups, t(82) = 1.290, p = .201. As shown in Figure 7a, the Fixed Pretesting group had a mean score of M=81.17 (SD=18.17), while the AI-Assisted Pretesting group had a mean score of M=76.27 (SD=16.48). These findings suggest that participants in both groups had a similar level of understanding of the regression analysis concepts covered in the recall test. This indicates that prior knowledge was unlikely to significantly impact the results of the subsequent activities and final test related to regression analysis.

\begin{figure}[h!]
    \centering
    \vspace{0 pt}
    \includegraphics[width=\textwidth]{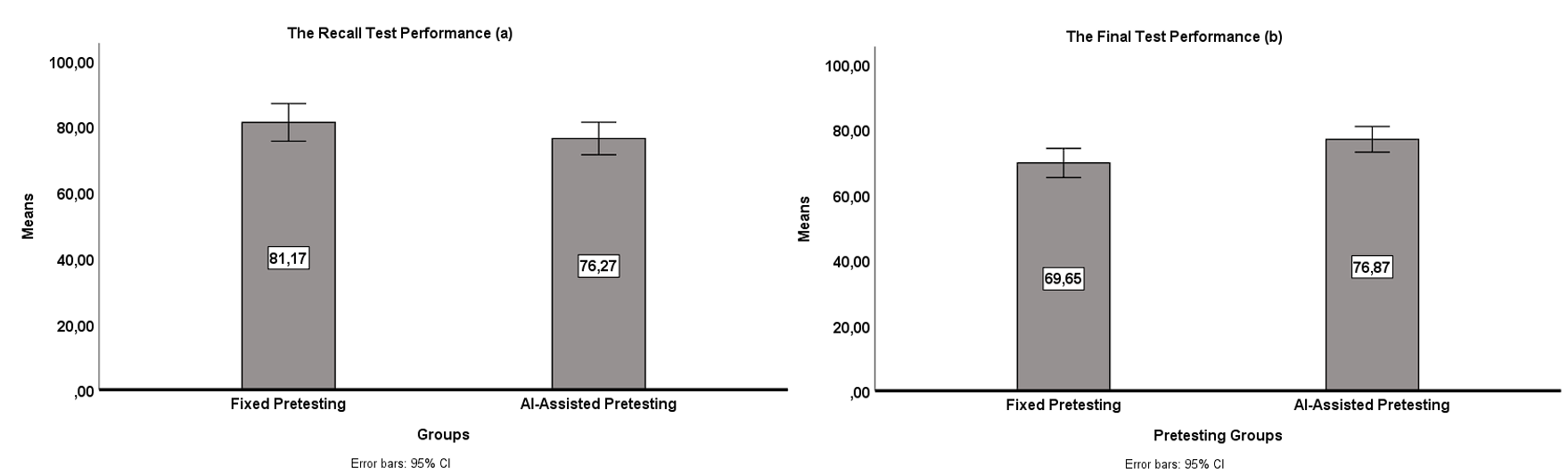} % Specify the image file here
    \caption{The Recall and Final Test Performances for Fixed Pretesting and AI-Assisted Pretesting groups (Study 3)} % Caption for the figure
    \label{fig:figure12} % Label for referencing in text
    \vspace{0pt}
\end{figure}

\textbf{Final test scores and motivational components} were analyzed using Multivariate Analysis of Variance (MANOVA), as this method allows for the evaluation of two or more continuous dependent variables (e.g., final test scores and motivation) in relation to a single categorical independent variable (group: FP or AP). The assumption of homogeneity of covariance matrices was tested using Box’s M Test to ensure that the data met the necessary conditions for MANOVA.

\subsection{Results}
The analysis examined the effects of group assignment (control vs. experimental) on six dependent variables: Interest and Enjoyment, Tension and Pressure, Perceived Choice, Perceived Competence, Perceived Value, and Final Test Scores. The independent variable, Group, had two levels: control and experimental.

Descriptive statistics for each dependent variable are provided in Table 2. The experimental group consistently outperformed the control group across most variables, particularly in Final Test Scores and the motivational dimensions of Interest and Enjoyment, Perceived Choice, and Perceived Competence. Additionally, the experimental group reported slightly lower scores in Tension and Pressure, a negative construct, indicating reduced pressure and tension during the learning process.

\begin{table}[ht]
\centering
\vspace{0pt}
\caption{Descriptive Statistics for Dependent Variables}
\begin{tabular}{lcccc}
\hline
\textbf{Dependent Variable} & \multicolumn{2}{c}{\textbf{Control}} & \multicolumn{2}{c}{\textbf{Experiment}} \\
 & \textbf{M} & \textbf{SD} & \textbf{M} & \textbf{SD} \\
\hline
Interest Enjoyment & 29.08 & 11.8 & 33.35 & 7.0 \\
Tension Pressure & 15.39 & 8.4 & 14.22 & 7.72 \\
Perceived Choice & 15.84 & 4.86 & 17.43 & 3.57 \\
Perceived Competence & 22.68 & 6.4 & 24.37 & 4.9 \\
Perceived Value & 19.87 & 6.25 & 20.9 & 4.19 \\
The Final Test Performance & 69.65 & 17.03 & 76.87 & 10.72 \\
\hline
\multicolumn{5}{l}{\textit{Note.} N = 87 (Control = 38, Experimental = 49)} \\
\end{tabular}
\label{tab:descriptive_statistics}
\vspace{0pt}
\end{table}

Box’s M test for homogeneity of covariance matrices was significant (Box’s M = 36.498, F= 1.603, p= 0.039); however, following the recommendations of Tabachnick et al. \cite{tabachnick2013using}, p-values above 0.001 are not considered a violation of this assumption. Thus, the test result was deemed non-problematic for the analysis.

The multivariate analysis of variance (MANOVA) revealed a marginally significant overall effect of Group on the dependent variables (Wilks’ Lambda, 
$\Lambda$=0.869,F(6,80)=2.019, p=0.073, partial $\eta^2_p$=0.131), accounting for 13.1\% of the total variance. While the multivariate effect was not statistically significant, medium effect sizes suggested a meaningful influence of group assignment.

Subsequent between-subjects analyses were conducted to investigate group differences for each dependent variable individually. These results are presented in Table 3. Significant differences were observed for Final Test Scores and Interest and Enjoyment. The results indicated that the experimental group reported significantly higher levels of Interest and Enjoyment (F(1,85)= 4.417, p=0.039, partial $\eta^2_p$=0.049) compared to the control group. Additionally, the experimental group scored significantly higher on Final Test Performance (F(1,85)= 5.839, p=0.018, partial $\eta^2_p$=0.064). As illustrated in Figure 7b, the Fixed Pretesting group had a mean score of M=69.65 (SD:17.03), while the AI-Assisted Pretesting group scored M=76.87 (SD=10.72).

\begin{table}[ht]
\small
\centering
\caption{Between-Subjects Results for Dependent Variables}
\resizebox{\textwidth}{!}{
\begin{tabular}{lcccccc}
\hline
\textbf{Dependent Variable} & \textbf{Type III Sum} & \textbf{df} & \textbf{Mean} & \textbf{F} & \textbf{Sig.} & \textbf{Partial} \\
 & \textbf{of Squares} & & \textbf{Square} & & & \textbf{$\eta^2$} \\
\hline
Interest Enjoyment & 389.859 & 1 & 389.859 & 4.417 & 0.039 & 0.049 \\
Tension Pressure & 29.31 & 1 & 29.31 & 0.455 & 0.502 & 0.005 \\
Perceived Choice & 53.867 & 1 & 53.867 & 3.087 & 0.083 & 0.035 \\
Perceived Competence & 60.632 & 1 & 60.632 & 1.932 & 0.168 & 0.022 \\
Perceived Value & 22.685 & 1 & 22.685 & 0.842 & 0.361 & 0.010 \\
The Final Test Performance & 1116.170 & 1 & 1116.170 & 5.839 & 0.018 & 0.064 \\
\hline
\end{tabular}
}

\label{tab:between_subjects_results}
\vspace{0pt}
\end{table}

The effect sizes for these variables indicated that 3.9\% to 6.4\% of the variance was explained by group assignment. No statistically significant differences were found for Tension and Pressure, Perceived Choice, Perceived Competence, or Perceived Value. Other dependent variables, while not significantly different, showed trends consistent with the benefits of the intervention.
%These results suggest that the intervention, which included AI-assisted pretesting, was effective in enhancing Interest and Enjoyment and improving performance on the final test. This provides evidence for the practical utility of the AI-assisted pretesting strategy in fostering motivation and learning outcomes.
\subsection{Discussion}
Study 3 demonstrates that the nature of the task significantly influences the effectiveness of AI-assisted pretesting strategies. In the higher-order task requiring iterative reasoning and decision-making, participants who used the adaptive AI Coach not only performed better on the final test but also reported higher interest and enjoyment. These findings highlight (a) the potential of AI-assisted pretesting to enhance students’ motivation in learning environments by fostering a more dynamic and engaging experience and (b) the need to align pretesting approaches with the cognitive demands of the task.

\section{General Discussion}
The purpose of our study was to investigate the efficacy of pretesting as a learning strategy when using generative AI tools like ChatGPT. Across three studies, we examined whether pretesting improves learning outcomes and explored how variations in pretesting format and task type influence its effectiveness. 

Study 1 confirmed that pretesting before AI assistance fosters retention of statistical knowledge. Study 2 explored whether AI-assisted pretesting, which dynamically adjusts questions based on student responses, provides advantages over traditional pretesting with fixed, pre-written questions. The results revealed no significant differences, suggesting that AI adaptation may not be as beneficial for computational tasks, where problem-solving follows structured and predictable steps. When learners engage with tasks that primarily involve recalling formulas or performing straightforward calculations, the dynamic nature of AI-assisted pretesting may not provide a substantial learning advantage over pre-written pretesting questions. Study 3 further investigated whether task complexity moderates pretesting effectiveness. Here, AI-assisted, adaptive pretesting significantly enhanced both motivation and learning outcomes, particularly for complex tasks that require iterative reasoning, decision making and conceptual understanding. By dynamically tailoring pretesting questions, AI can better support students in navigating open-ended problems and encourage deeper cognitive engagement, a benefit not observed in more structured computational tasks.

One critical consideration in these findings is the role of productive struggle in learning. Research suggests that engaging with challenging tasks and struggling to generate solutions can lead to deeper understanding and long-term retention \cite{warshauer2015productive, hiebert2007effects}. While GenAI tools provide immediate access to solutions, this convenience may reduce the opportunities for productive struggle. Our findings support the idea that pretesting serves as a structured mechanism for preserving productive struggle in AI-assisted learning environments. By requiring students to engage with a problem before accessing ChatGPT, pretesting encourages them to activate prior knowledge and think critically about potential solutions. 

Although AI-assisted pretesting results in deeper cognitive engagement, students may not always recognize its benefits. Research on human-AI interactions has shown that users' perceptions of AI helpfulness do not always align with actual learning gains. In some cases, AI systems that provide quick and confident responses are rated as more helpful, even when they do not lead to better task performance or deeper understanding \cite{lee2022evaluating}. In our work, while adaptive pretesting led to improved performance (Study 3), not all motivational measures reflected this benefit. Specifically, differences in Perceived Choice and Perceived Value between the adaptive and fixed pretesting groups were not significant. This suggests that although adaptive pretesting results in better learning performance, students may not always recognize its value. Future research should explore strategies to help learners better understand the role of productive struggle and the benefits of delayed feedback in AI-assisted learning environments, ensuring that students do not equate ease of access to information with effective learning.

Together, our three studies contribute to a growing body of research demonstrating the benefits of pretesting before accessing to-be-learned information \cite{grimaldi2012and, kornell2009unsuccessful, kornell2014attempting, richland2009pretesting}.  Our findings extend this line of research into the context of GenAI tools, highlighting their potential to complement pretesting strategies in fostering better retention. While several studies have reported limited benefits of pretesting, especially in classroom settings \cite{carpenter2018effects}, our results show that when integrated with GenAI, pretesting can produce significant improvements in learners' ability to retain and apply complex concepts \cite{giebl2021answer,giebl2023thinking}.

The current findings align with prior research suggesting that reliance on the Internet for information retrieval reduces internal knowledge retention\cite{marsh2019digital, sparrow2011google}. As people increasingly depend on digital tools for accessing information, they may encode less factual knowledge internally and instead develop a "memory partnership" with search-based AI technologies \cite{sparrow2011google}. This shift in cognitive habits means learners may prioritize remembering how to access information rather than internalizing it. However, our findings suggest that pretesting before engaging with GenAI tools may help mitigate this tendency. In line with Giebl et al. \cite{giebl2021answer}, we propose that pretesting can serve as an effective cognitive scaffold that enables  students to actively engage with key concepts before relying on AI for answers.

%One explanation for the pretesting effect relates to its role in helping learners accurately assess their own knowledge (see \cite{giebl2021answer}). Research indicates that access to web search tools often inflates individuals' confidence in their cognitive abilities, resulting in an overestimation of what they know \cite{ward2013one, hamilton2018blurring}. This phenomenon, known as cognitive self-esteem (CSE), reflects an individual’s belief in their ability to think and recall information \cite{ward2013one}. Studies have shown that individuals who use Google to answer trivia questions report significantly higher CSE scores than those without access \cite{ward2013one}. Similarly, Hamilton and Yao \cite{hamilton2018blurring} found higher CSE scores for those who used Google than for those in the no-Google or control conditions. Our findings suggest that incorporating pretesting with GenAI tools like ChatGPT may help learners calibrate their confidence more accurately, ensuring they engage with their internal cognitive resources before seeking external assistance.

Beyond improving knowledge retention, our findings provide insights into how AI might be integrated into educational settings to support learning. While AI systems are often designed to maximize efficiency, our results indicate that AI-driven learning tools have the potential to balance assistance with cognitive challenge. Rather than replacing struggle, AI could be leveraged to enhance it—offering the right level of scaffolding at key moments to foster deeper engagement. Adaptive pretesting appears to be one such approach, as it encourages students to engage with key concepts before receiving AI-generated explanations. Future research could explore additional mechanisms for strategically delaying assistance, allowing students to experience struggle and develop problem-solving strategies before AI intervention.

Overall, these studies highlight the potential for AI-assisted pretesting to enhance learning outcomes. While traditional pretesting strategies remain valuable, AI-driven adaptive pretesting appears to offer unique advantages when tasks require deeper engagement and iterative reasoning. By demonstrating that pretesting remains an effective learning strategy in AI-supported environments, our findings underscore the importance of structured cognitive engagement before interacting with generative AI. As AI tools become increasingly integrated into education, ensuring that students engage meaningfully with learning materials before turning to AI assistance will be critical for fostering deep, lasting learning.

While our results offer evidence-based insights into the benefits of pretesting, they also highlight certain constraints. AI is a powerful tool for expanding access to education, but it must be carefully designed to support meaningful learning rather than shortcutting cognitive effort. 

\subsection{Future Research Directions}
Building on the insights gained, future research can focus on several key areas. First, tailoring AI-assisted pretesting strategies to individual learner profiles could significantly enhance their effectiveness. Considering variations in students’ prior knowledge when designing pretesting approaches may help maximize learning outcomes and ensure that these strategies are more effectively adapted to individual needs. Second,  our work suggests that AI-assisted pretesting appears most beneficial for complex, cognitively demanding tasks. However, “complexity” can vary greatly—ranging from straightforward single-step questions to multi-stage, conceptually integrated problems. Thus, future investigations should explore how learning gains differ when students encounter AI-assisted pretesting at varying levels of task complexity. Such insights will help better match pretesting difficulty with learners’ readiness. Finally, understanding how AI-assisted pretesting influences long-term memory and practical application can provide deeper insights into the pedagogical benefits of this approach and guide the development of more effective educational practices.

\subsection{Limitations} While our work shed light on the potential benefits of AI-assisted pretesting, it has several limitations. First, the sample was drawn from a single institution and focused primarily on an intermediate-level statistics course, which may constrain the generalizability of our results to other settings and subject areas. Second, the tasks used reflect only a subset of possible complexities and may not capture the full range of challenges learners face in diverse domains. Third, our study did not measure long-term retention, leaving open the question of how pretesting interventions influence sustained mastery over time. Addressing these issues in future research will provide a more comprehensive understanding of AI-assisted pretesting’s effectiveness and scalability.

\section{Conclusion} Taken together, these findings advance our understanding of how classic learning strategies can be effective in the context of modern educational technology. First and foremost, we have learned that pretesting remains a potent learning enhancer in the age of AI. This directly corroborates and extends prior literature on the pretesting effect. As Pan and Carpenter’s review \cite{pan2023prequestioning} emphasizes, decades of empirical work have shown pre-instruction testing can often improve learning outcomes. Our study adds new evidence that this holds true when learners engage with generative AI as part of the study phase. It suggests that the underlying cognitive mechanisms identified in earlier research (e.g. enhanced attention, error-based learning, activation of retrieval routes) continue to operate in an AI-driven learning environment. This is a notable theoretical contribution. In essence, our work shows that the synergy of human memory activation and computer-provided knowledge can produce superior learning – a result that educators and technologists can leverage in designing future tools and curricula.

Moreover, our findings contribute new insights about when and how technology enhances the pretesting effect. A nuanced understanding is that the pedagogical payoff of integrating AI with pretesting is most evident on complex tasks that demand reasoning. AI’s impact is maximized when learners are first challenged to generate ideas or answers on their own. In simpler terms, the computer is most useful as a learning partner after the student has tried to solve the problem themselves. This resonates with constructivist theories and the concept of “productive failure” in education, where initial struggle leads to better understanding \cite{kapur2010productive}. By empirically showing that AI-assisted pretesting leads to superior retention of complex concepts, we provide evidence that the right integration of computers can amplify a learning strategy that previously yielded mixed results in certain settings. For example, Carpenter et al. \cite{carpenter2018effects} found relatively limited gains from prequestions in a conventional classroom scenario, but our results demonstrate that pairing prequestions with an interactive AI changes the equation – producing substantial improvements in students’ ability to retain and apply challenging material. This suggests that the effectiveness of pretesting is not fixed, but can be elevated by tools that engage learners in adaptive, responsive ways.

Our study also sheds light on the ongoing debate about the role of computers as cognitive partners in learning. Prior works have raised concerns that heavy reliance on computers (e.g. search engines or AI) might weaken individuals’ own knowledge retention – the so-called “Google effect,” where people remember how to find information rather than the information itself \cite{sparrow2011google}. Our evidence offers a more optimistic perspective by identifying a strategy to counteract this effect. We found that inserting a thoughtful pretesting phase before students turn to AI essentially forces them to think about key concepts upfront, which in turn mitigates the tendency to offload all mental work to the machine.

\bibliographystyle{unsrt}  


\bibliography{references}
\end{document}